\documentstyle[12pt,epsf]{article}

\newlength{\pubnumber} 

\catcode`\@=11
\@addtoreset{equation}{section}

\def\section{\@startsection{section}{1}{\z@}{3.5ex plus 1ex minus .2ex}
 {2.3ex plus .2ex}{\large\bf}}
\def\subsection{\@startsection{subsection}{2}{\z@}{2.3ex plus .2ex}
 {2.3ex plus .2ex}{\bf}}

\begin{document}

\begin{titlepage}
\samepage{
\setcounter{page}{1}
\rightline{ACT-2/00}
\rightline{BU-HEPP-02/14}
\rightline{CASPER-02/06}
\rightline{CTP-TAMU-41/98}
\rightline{OHSTPY-HEP-T-98-023}
\rightline{\tt hep-th/0002089}
\rightline{August 2002}
\vfill
\begin{center}
 {\Large \bf Ratio of Quark Masses in Duality Theories}
\vskip 2.0truecm
       {\large
        G.\  Cleaver,$^{1,2,3}$\footnote{Gerald\underline{\phantom{C}}Cleaver@baylor.edu}
        and
        K.\  Tanaka,$^4$\footnote{tanaka@mps.ohio-state.edu}}
\\
\vspace{.12in}
{\it $^{1}$ Center for Astrophysics, Space Physics \& Engineering Research\\
            Dept.\  of Physics, PO Box 97316, Baylor University,\\
            Waco, TX , 76798-7316USA\\}
\vspace{.06in}
{\it $^{2}$ Center for Theoretical Physics,
            Dept.\  of Physics, Texas A\&M University,\\
            College Station, TX 77843, USA\\}
\vspace{.06in}
{\it $^{3}$ Astro Particle Physics Group,
            Houston Advanced Research Center (HARC),\\
            The Mitchell Campus,
            Woodlands, TX 77381, USA\\}
\vspace{.06in}
{\it$^{4}$  Department of Physics, The Ohio State University, 
            Columbus, OH 43210-1106, USA\\}
\vspace{.025in}
\end{center}
\vskip 2.0truecm
\begin{abstract}
We consider $N=2$ $SU(2)$ Seiberg--Witten duality theory for 
models with $N_f=2$ and $N_f=3$ quark flavors.
We investigate arbitrary large bare mass ratios
between the two or three quarks at the singular points. 
For $N_f=2$ we explore large bare mass ratios
corresponding to a singularity in the strong coupling region.
For $N_f=3$ we determine the location of both strong and weak
coupling singularities that produce specific large bare mass ratios. 
\end{abstract}
\smallskip}
\vfill
\end{titlepage}

\setcounter{footnote}{0} 

\def\at{ }
\def\beq{\begin{equation}}
\def\eeq{\end{equation}}
\def\beqn{\begin{eqnarray}}
\def\eeqn{\end{eqnarray}}
\def\no{\noindent }
\def\nolabel{\nonumber }

\def\NA{non-Abelian }

\def\app{\approx}
\def\gsim{{\buildrel >\over \sim}}
\def\lsim{{\buildrel <\over \sim}}
\def\lt{<}

\def\ie{i.e., }
\def\eg{{\it e.g.}}
\def\eq#1{Eq.\ (\ref{#1})}
\def\etal{{\it et al\/}}

\def\slash#1{#1\hskip-6pt/\hskip6pt}
\def\slk{\slash{k}}

\def\dag{\dagger}
\def\qandq{\quad {\rm and} \quad} 
\def\qand{\quad {\rm and} } 
\def\andq{ {\rm and} \quad } 
\def\qwithq{\quad {\rm with} \quad} 
\def\qwith{ \quad {\rm with} } 
\def\withq{ {\rm with} \quad} 

\def\fhalf{\frac{1}{2}}
\def\fsqrt{\frac{1}{\sqrt{2}}}
\def\half{{\textstyle{1\over 2}}}
\def\third{{\textstyle {1\over3}}}
\def\fourth{{\textstyle {1\over4}}}
\def\quarter{{\textstyle {1\over4}}}
\def\sixth{{\textstyle {1\over6}}}
\def\nineth{{\textstyle {1\over9}}}
\def\m{$\phantom{-}$}
\def\j{$-$}
\def\ps{{\tt +}}
\def\pps{\phantom{+}}

\def\zz{$Z_2\times Z_2$ }

\def\Tr{{\rm Tr}\, }
\def\tr{{\rm tr}\, }

\def\MP{M_{P}}
\def\GeV{\,{\rm GeV}}
\def\TeV{\,{\rm TeV}}

\def\lam#1{\lambda_{#1}}
\def\non{\nonumber}
\def\SM{Standard--Model }
\def\SUSY{supersymmetry }
\def\SSSM{supersymmetric standard model}
\def\MSSM{minimal supersymmetric standard model}
\def\MSSSM{MS$_{str}$SM }
\def\MSSSMc{MS$_{str}$SM, }
\def\obs{{\rm observable}}
\def\sig{{\rm singlets}}
\def\hid{{\rm hidden}}
\def\MS{M_{str}}
\def\Ms{$M_{str}$}

\def\vev#1{\langle #1\rangle}
\def\mvev#1{|\langle #1\rangle|^2}

\def\y{\,{\rm y}}
\def\l{\langle}
\def\r{\rangle}
\def\o#1{\frac{1}{#1}}

\def\zi{z_{\infty}}

\def\calF{{\cal F}}
\def\smallmatrix#1#2#3#4{{ {{#1}~{#2}\choose{#3}~{#4}} }}
\def\Minv{{ (M^{-1}_\ab)_{ij} }}
\def\ii{{(i)}}
\def\eps{\epsilon}

\def\disc{discriminant\ }

\def\inbar{\,\vrule height1.5ex width.4pt depth0pt}

\def\IC{\relax\hbox{$\inbar\kern-.3em{\rm C}$}}
\def\IQ{\relax\hbox{$\inbar\kern-.3em{\rm Q}$}}
\def\IR{\relax{\rm I\kern-.18em R}}
 \font\cmss=cmss10 \font\cmsss=cmss10 at 7pt
 \font\cmsst=cmss10 at 9pt
 \font\cmssn=cmss9

\def\IZ{\relax\ifmmode\mathchoice
 {\hbox{\cmss Z\kern-.4em Z}}{\hbox{\cmss Z\kern-.4em Z}}
 {\lower.9pt\hbox{\cmsss Z\kern-.4em Z}}
 {\lower1.2pt\hbox{\cmsss Z\kern-.4em Z}}\else{\cmss Z\kern-.4em Z}\fi}

\hyphenation{su-per-sym-met-ric non-su-per-sym-met-ric}
\hyphenation{space-time-super-sym-met-ric}
\hyphenation{mod-u-lar mod-u-lar--in-var-i-ant}
\def\mbf{\mathbf}
\def\L{\Lambda}
\def\Lt{\Lambda^2}

\section{$\mbf N=2$ $\mbf SU(2)$ QCD with Bare Mass Quarks} 

Understanding of the vacuum structure of $N=2$ supersymmetric gauge theories in
four spacetime dimensions has progressed significantly in recent years. 
For example, the moduli space of $N=2$ supersymmetric 
$SU(2)$ QCD is now known to be the complex $u$--plane with its singularities. 
Physically, $u$ is the vacuum expectation value of the square of 
a complex scalar field, $\phi$, 
in the adjoint representation of $SU(2)$,
$u= \vev{{\rm Tr}\, \phi^2}$. 
The $u$--plane singularities 
are described by their monodromy matrices \cite{sw1,sw2}. 
To every value of $u$ 
there corresponds a genus one Riemann surface that can be represented by 
a curve of the form 
\beqn
y^2 = F(x,u)\, ,  
\label{eqy}
\eeqn 
where $F$ is a cubic polynomial in $x$,
\beqn
F = x^3 + \beta x^2 + \gamma x + \delta\, .
\label{defcub}
\eeqn
Thus, (\ref{eqy}) yields a
family of elliptic curves over the parameter space of $u$.
 
Associated with any polynomial $F$ is its
discriminant  $\Delta$ defined by
\beqn
\Delta = \prod_{i<j} (e_i - e_j)^2,
\label{defdisc}
\eeqn
where the $e_i$ are the roots of $F$. 
The branch points of the $N=2$ family of curves $y^2 = F(x,u)$ 
overlap at the locations where the discriminant $\Delta$ is zero.  
In other words, the zeros of $\Delta$
specify the locations of the singularities in $u$ parameter space. 
At the singularities, certain magnetic monopoles or dyons
become massless.
For a cubic polynomial $F$ (\ref{defcub}),
the discriminant (\ref{defdisc})
can be expressed as
\beqn
\Delta = -27 \delta^2 + 18 \beta \gamma \delta 
        + \beta^2 \gamma^2 - 4 \beta^3 \delta - 4 \gamma^3\, .
\label{disccub}
\eeqn

In this letter, we examine the relationship between bare mass ratios
of quark flavors and the location of the singularities. While the 
bare mass ratios are, indeed, free parameters of the theory,  
we show that the discriminant has predictive capabilities with 
regard to bare quark mass ratios {\it at the singularities}. 
We investigate this for both the two flavor and the three flavor cases.

\section{The Two Quark Model}

Consider first the $N=2$, $N_f=2$ Seiberg--Witten $SU(2)$ model, 
with related non--zero bare masses denoted $m_a$ and $m_b$, where $m_a \ge m_b$. 
The family of curves of the modular space can be parametized by
\beqn 
  y^2 = (x^2 - t^2 ) (x - u) +   2 m^2 t x - 2 M^2 t^2\, ,
\label{fce2m}
\eeqn
where the square of the energy scale of the theory is
$t\equiv \frac{1}{8}\Lambda^2$. We have also used    
$m^2\equiv m_a m_b$ and 
$M^2\equiv \half (m_a^2 + m_b^2)$.
This equation is derived by
Seiberg and Witten on the basis of conservation laws and appropriate
boundary conditions \cite{sw1,sw2}.
In (\ref{fce2m}), $x$, $u$, and $t$ have mass dimension 2, 
while $m$, $M$, and $N$ all have mass dimension 1. 
The mass dimension is one--half of the $U(1)_R$ charge of \cite{sw1,sw2}. 

Let us examine the possible mass hierarchy between
the bare masses $m_a$ and $m_b$. When the masses are equal,
the \disc of (\ref{fce2m}) is
\beqn
\Delta = 4 t^2 [ (u + t)^2 - 8 m^2 t ] (u - t - m^2)^2 \, .
\label{de2ma} 
\eeqn
When the masses are unequal, let 
\beqn
M^2 &=&   m^2 + (M^2 - m^2) \equiv    m^2 + 2 D^2
\label{mddefa}
\eeqn
and
\beqn
M^2 &=& - m^2 + (M^2 + m^2) \equiv  - m^2 + 2 N^2. 
\label{mndefa}
\eeqn
In other words,
\beqn
D^2 &=& \half (M^2 - m^2) = \fourth (m_a - m_b)^2\, ,
\label{ddefa}
\eeqn
and
\beqn
N^2 &=& \half (M^2 + m^2) = \fourth (m_a + m_b)^2\, .\label{ndefa}
\eeqn
This gives
\beqn
\Delta = 4 t^2 \left[ (u + t)^2 - 8 m^2 t \right] (u - t - m^2)^2 + 
\Delta_{BNDY}\, ,
\label{de2mb} 
\eeqn
where
\beqn
\Delta_{BNDY} = - 144 D^2 t^2 \left[ (3 t^2 - t u) N^2 + t u D^2 - t^2 u + 
\nineth u^3\right]\, .  
\label{de2mabndy} 
\eeqn
The condition $D^2=0$ yields,  by definition, the equal mass case $m_a = m_b$;
$D^2>0$ implies  $m_a > m_b$.
 
We kept the form of the first term of $\Delta$ in (\ref{de2mb}) 
similar to that in (\ref{de2ma}) so  we can decide on the region in 
the $u$ space that we wish to focus on.  
At or near the singularities in the $u$--plane,
\beqn
\Delta_{BNDY}=0
\label{eqbndy}
\eeqn 
can be viewed as a boundary constraint 
when $m_a$ and $m_b$ are inequivalent.
For $D^2\ne 0$, (\ref{eqbndy}) implies
\beqn
(3 t^2 - t u) N^2 +  t u D^2 - t^2 u + \nineth u^3 = 0\, .
\label{ndbdry}
\eeqn

The three distinct roots of $\Delta$ in (\ref{de2ma}) 
(and $\Delta$ in (\ref{de2mb}) when $\Delta_{BNDY} = 0$) are
$u = u_o\equiv t + m^2$,  $u= u_{+}\equiv -t + \sqrt{8 m^2 t}$,
and $u= u_{-}\equiv -t - \sqrt{8 m^2 t}$.
Consider the singular region around the double zero of $\Delta$,
$u_o \equiv t + m^2$.  At this singularity, we find 
\beqn
m^2 &=& u - t = N^2 - D^2 
\label{d2defda}.
\eeqn
Together (\ref{d2defda}) and (\ref{ndbdry}) yield,
\beqn
N^2 &=& u \left( \frac{u}{3 t} - \frac{1}{27}\frac{u^2}{t^2}\right)
\label{n2sol}
\eeqn
and
\beqn
D^2 &=& t \left( 1 - \frac{u}{t} + \frac{u^2}{3 t^2} 
                   - \frac{1}{27}\frac{u^3}{t^3}\right)\, .
\label{d2sol}
\eeqn
For 
\beqn
\frac{u}{t}= 1 + \epsilon, 
\label{ueps}
\eeqn
where $\epsilon$ is regarded 
as small and positive, we obtain
\beqn
N^2 &=& t \frac{8}{27}\left(1 + \frac{15}{8}\epsilon\right)\, ,
\label{n2solb}
\eeqn
and
\beqn
D^2 &=& t \frac{8}{27}\left(1 - \frac{3}{2}\epsilon\right)\, .
\label{d2solb}
\eeqn

Expressed in terms of $N$ and $D$, the mass ratio is
\beqn
\frac{m_a}{m_b} = \frac{N+D}{N-D}\, .
\label{mrab}
\eeqn
From Eqs.\ (\ref{n2solb}-\ref{mrab}) we find
\beqn
\frac{m_a}{m_b}= \frac{N+D}{N-D} = \frac{32}{27\epsilon}
\label{mambr}
\eeqn
at the double zero singularity, $u= t + m^2$.
Since $\epsilon$ can be an arbitrarily small number as 
$u\rightarrow t$, the mass ratio can be arbitrarily large \cite{kt1}.

In concluding our study of the two mass case, we comment that
in \cite{ho} Hanany and Oz started with a hyperelliptic modular
curve
\beqn
y^2 = (x^2 - u + t)^2 - 64 t^2 (x + m_a)(x + m_b)\, ,
\label{hoeq}
\eeqn
and obtained the same $\Delta$ as in (\ref{de2ma})
for equal masses. For classical Lie groups, the Seiberg--Witten
curves may always be expressed as hyperelliptic curves. $SU(2)$
is the only classical group that also allows the corresponding 
Seiberg--Witten curve to take elliptic form \cite{oh}. 

\section{The Three Quark Model}

Now consider the $N=2$, $N_f=3$ Seiberg--Witten $SU(2)$ 
model with related non--zero bare masses, 
$m_a$, $m_b$, and $m_c$, where $m_a \ge m_b \ge m_c$.   
For three quarks, the family of curves equation is \cite{sw2}
\beqn 
  y^2 = x^2 (x - u) - t^2 (x - u)^2 - 3 M^2 t^2 (x - u) 
        +   2 m^3 t x - 3 P^4 t^2\, .
\label{fce3m}
\eeqn
Here we have defined
$t \equiv \Lambda/8$, 
$m^3 \equiv  m_a m_b m_c$,
$M^2 \equiv (m_a^2 + m_b^2 + m_c^2)/3$, and
$P^4 \equiv (m_a^2 m_b^2 + m_b^2 m_c^2 + m_c^2 m_a^2)/3$.
Note that $x$ and $u$ have mass dimension 2, 
while $t$, $m$, $M$, and $P$ all have mass dimension 1. 
Let us also define for later use the variables $G$ and $H$ via
\beqn
M^2 &=& m^2 + (M^2 - m^2) = m^2 + G^2\, ,\label{d2}
\eeqn
and
\beqn
P^4 &=& m^4 + (P^4 - m^4) = m^4 + H^4.\label{e4} 
\eeqn
In the case of  three equivalent bare masses, $m_a = m_b = m_c $,
we reach the limits $m = M = P$ and $G = H = 0$. 

Eq.\ (\ref{fce3m}) can be rewritten into the polynomial form of (\ref{defcub}),
with
\beqn
\beta  &=& - t^2 - u,\label{bv1}\\
\gamma &=&  2 t^2 u + 2 m^3 t - 3 M^2 t^2,\label{cv1}\\
\delta &=& - t^2 u^2 + 3 M^2 t^2 u - 3 P^4 t^2\, .\label{dv1}
\eeqn
From (\ref{disccub}), we find the corresponding discriminant to be
\beqn
\Delta &=& 
t^2 \left[-32 m^9 t - 243 P^8 t^2 - 6 P^4 (t^2 - 2 u) 
(27 M^2 t^2 + 2 t^4 - 8 t^2 u - u^2)\right.\nolabel\\
       & &
\phantom{t^2 1}
+ (12 M^2 + t^2 - 4 u)  (-3 M^2 t^2 + u^2)^2 
+ 4 m^6 (36 M^2 t^2 + t^4 - 22 t^2 u + u^2) \nolabel\\
       & &
\phantom{t^2 1}
- 4 m^3 t \left(54 M^4 t^2 - 27 P^4 (t^2 + u) 
+ u (-2 t^4 + 11 t^2 u - 11 u^2)\right. + 
\nolabel\\
       & & 
\phantom{t^2 1 - 4 m^3 t 1}
\left. \left. 3 M^2 (t^4 - 13 t^2 u + 10 u^2)\right)\right]
\label{discnm}
\eeqn

While the variables $m$, $M$, and $N$ simplify the form of the
discriminant (\ref{discnm}), the alternative set given above,
$m$, $G$, and $H$, are more useful for our mass ratio study. 
In the language of $G$ and $H$, 
the discriminant separates into four components: 
\beqn
\Delta = \Delta(m,u,t) + \Delta(m,G,u,t) + \Delta(m,H,u,t) +
          \Delta(m,G,H,u,t).                    
\label{discdiv}
\eeqn
The first component,
\beqn
\Delta(m,u,t) &=&
-t^2 (m^2 + m t - u)^3 \times
\nolabel\\ 
              & &
\left[
(32 m^3 t + 3 m^2 t^2 + 3 m t^3) + (t^2 - 12 m t) u - 4 u^2\right]
\label{delta}
\eeqn
contains only $m$, $u$, and $t$. The entire discriminant reduces
to just this term for the equal mass case $m_a = m_b = m_c$.
The second, third, and fourth terms,
\beqn
\Delta(m,G,u,t) &=&  
  3 G^2 t^2 \left[ 36 G^4 t^4 
+ 3 G^2 t^2 (-24 m^3 t + 36 m^2 t^2 +  t^4 + 4 t^2 u + 8 u^2)\right.
\nolabel\\& &\phantom{3 G^2 t^2 L} 
+          \left(48 m^6 t^2 - 144 m^5 t^3 + 54 m^4 (t^4 + 2 t^2 u)\right. 
\nolabel\\& &\phantom{3 G^2 t^2 L}
-    2 u^2 (t^4 - 4 t^2 u - 2 u^2) - 4 m^3 (t^5 - 13 t^3 u + 10 t u^2)  
\nolabel\\& &\phantom{3 G^2 t^2 L}
+ \left.\left.   6 m^2 (t^6 - 4 t^4 u - 8 t^2 u^2)\right) \right]
\label{deltaD}\\
\Delta(m,H,u,t) &=& 
 -H^4 t^2 \left[ 243 H^4 t^2  + 6 \left( 81 m^4 t^2 
+(  27 m^2 t^2 + 2 t^4 - 8 t^2 u -  u^2 )\times \right.\right. 
\nolabel\\& & \phantom{-H^4 t^2 L}\left.\left. (t^2 - 2 u) 
- 18 m^3 t (t^2 + u)\right)\right]
\label{deltaE}\\
\Delta(m,G,H,u,t) &=& - 162 G^2 H^4 t^4 (t^2 - 2 u) 
\label{deltaDE}
\eeqn
additionally contain, $G$, $H$, and both $G$ and $H$, respectively. 

Moving away from the equivalent mass point, we can still
effectively keep
\beqn
\Delta = \Delta(m,u,t)
\label{dm}
\eeqn
by separately enforcing an additional boundary constraint 
\beqn
\Delta_{BNDY}\equiv\Delta(m,G,u,t)+\Delta(m,H,u,t)+\Delta(m,G,H,u,t)=0\,\,.
\label{dmDE}
\eeqn
Imposition of the boundary constraint (\ref{dmDE}) allows us to 
solve for $u$ at the singular points simply in terms of $m$ and $t$.
That is, the singularities are located at the values of $u$ such that
$\Delta(m,u,t)= 0$.
There is a triple zero of (\ref{delta}) at
\beqn
u \equiv u_o= m^2 + m t 
\label{ztrip}
\eeqn
along with additional zeros at
\beqn
u &\equiv& u_+ = \frac{1}{8}
\left(-12 m t + t^2 + \sqrt{t (8 m + t)^3}\right) \,\, ,
\label{zpos}
\eeqn
and
\beqn
u &\equiv& u_- = \frac{1}{8}
\left(-12 m t + t^2 - \sqrt{t (8 m + t)^3}\right) \,\, .
\label{zneg}
\eeqn

In the $(\root 3\of{|m_a m_b m_c|} = m)<<t$ limit,  
\beqn
u_{o} &\rightarrow& m t = m \Lambda / 8 
\label{u0w}\\
u_{+} &\rightarrow& 2 t^2 = \Lambda^2/32
\label{upw}\\
u_{-} &\rightarrow& -3 m t = - 3 m \Lambda/8\, .
\label{umw}
\eeqn
Therefore for small $m$, we find $|u|\ll \Lambda^2 $
in the regions near any of the three singularities.
Since weak coupling corresponds to $|u|\gg \Lambda^2$,
$m\ll t$ implies 
strong coupling in the neighborhood of the singularities.
Strong coupling also  
results when $m$ and $t$ are of the same magnitude. 
Only in the $m>> t$
limit do the $u$ singularities move into the weak coupling realm.
 
At each of the three distinct zeros of (\ref{delta}), the
direct dependence of our boundary discriminant $\Delta_{BNDY}$ 
on $u$ is removed 
by making the appropriate root substitution, 
(\ref{ztrip}), (\ref{zpos}), or (\ref{zneg}). 
We can always scale $t$ to unity, thereby
effectively defining $m$, $G$, and $H$ in units of $t$.
Thus, at a given singularity, $\Delta_{BNDY}$ becomes
a polynomial involving only $m$, $G$, and $H$,  
\beqn
\Delta_{BNDY}\equiv \Delta(m,G) + \Delta(m,H) + \Delta(m,G,H) = 0\, .
\label{bdeq1}
\eeqn 
for $u= u_o$ or $u_{+}$ or $u_{-}$. 

We can solve (\ref{bdeq1}) for any variable from the
set $\{m,\, G,\, H\}$ in terms of the other two.  
Recall however that, irregardless of (\ref{bdeq1}),
$m$, $G$, and $H$ are not totally independent parameters, 
at least if they are to result in necessarily real and positive 
$m_a^2$, $m_b^2$, and $m_c^2$.  
Some $(m,\, G,\, H)$ solutions to (\ref{bdeq1}) 
may result in unacceptable
(i.e., negative or complex) values of the bare mass--squares.  

The three equations 
defining
$m$, $M$ and $P$ may be combined to form a polynomial 
\beqn
&&x^3 - 3 M^2 x^2 + 3 P^4 x - m^6 = 0,\label{polyx1}
\eeqn            
where 
$x\equiv m_a^2$ or $m_b^2$ or $m_c^2$.
Equivalently,
\beqn
&&x^3 - 3 (m^2 +G^2) x^2 + 3 (m^4 + H^4) x - m^6 = 0.
\label{polyx2}            
\eeqn
Thus, the viable $m$, $G$, and $H$ combinations  
are those such that {\it all three}
roots of (\ref{polyx2}), corresponding to $m_a^2$, $m_b^2$, and $m_c^2$,
are real and positive. One trivial constraint is $H,G\ge 0$.
Further, we find that $H=0$ is 
physically allowed only when $G=0$ simultaneously, i.e., when all
masses are equivalent. Specifically $H=0$ and $G> 0$
implies that two mass--squares are negative, and likewise for 
$G=0$ and $H>0$. 

One approach to generating a consistent set of masses 
$\{m_a,m_b,m_c\}$ for a given 
$u=u_o$ or $u_+$ or $u_-$ singularity
is to determine the general structure of 
$m$, $G$, and $H$ solutions to $\Delta_{BNDY}=0$. 
An alternate, albeit less general, method is to rewrite the boundary 
constraint (\ref{bdeq1}) directly in terms of the three bare masses, 
$m_{i=a,b,c}$. (See Appendix A.)
Following this, we can specify a ratio between the three masses,
\beqn
 m_a/m_a : m_b/m_a : m_c/m_a = 1: b: c\, ,
\label{abcratio}
\eeqn
where $1 \ge b = m_b/m_a \ge c = m_c/m_a > 0$. 
Next we choose a singularity type 
$u= u_o$ (\ref{ztrip}), $u=u_+$ (\ref{zpos}), or $u=u_-$ (\ref{zneg}),
and rewrite $u$ in terms of $m_a$, $m_b$ and $m_c$. 
We then substitute $ b\, m_a$ and $ c\, m_a$ 
for $m_b$ and $m_c$ in $\Delta_{BNDY}=0$.   
Hence, we can determine the allowed values of $m_a$ for the given
mass ratio (\ref{abcratio}). Knowledge of $m_a$, $m_b$, and $m_c$
and the singularity type specifies the location of the associated singularity.

After a mass ratio (\ref{abcratio}) is {\it chosen},  
the boundary constraint appears for the $u_o$ singularity 
as a polynomial of tenth order having at least four zero roots.
Thus, the non--trivial $m_a$ solutions are the roots of 
a sixth order polynomial.
At the $u_+$ and $u_-$ singularities the boundary constraint
appears as an eighth--order polynomial (with at least two zero roots),
with some terms generically containing an extra factor of $\sqrt{1 + r\,  m_a}$, 
where $r$ is a numerical coefficient.  
Roots of the sixth and eight order polynomials
can be found using programs such as Mathematica or Maple.  

We followed the mass ratio approach to learn where a $u$ singularity
is consistent with a large bare mass ratio.
We considered, for example, the location of the singularities when
the bare mass ratio 
is the order of the physical top, charm, and up mass ratio,  
$m_a/m_a:m_b/m_a:m_c/m_c
\app 1 : 7\times 10^{-3}: 3\times 10^{-5}$. 
For this ratio,
the $u_o$ singularity provides a solution of  
$m_a \app  1600 \Lambda$, $u= u_o \app 74 \Lambda^2$, which is 
still in the strong coupling region. 
The $u_+$ and $u_-$ singularities 
offer similar strong coupling solutions: 
$m_a \approx 1100 \Lambda$ at $u= u_+ \app +13  \Lambda^2$ and
$m_a \approx  820 \Lambda$ at $u= u_- \app -9.7 \Lambda^2$, respectively.
In Table I of Appendix B we also present examples of large bare   
mass ratios for $u_o$, wherein the $m_a$ and $u$ solutions are in the ranges
$0.1\, \Lambda   \lsim m_a \lsim 2000 \Lambda$ and
$0.1\, \Lambda^2 \lsim u   \lsim 200  \Lambda^2$.

Generic three quark bare mass ratios have $u_o$ solutions with   
$m_a\gsim \Lambda$ and $|u|\gsim \Lambda^2$.
For a given mass ratio,
the $u_+$ and $u_-$ singularities typically, but not always, offer  
legitimate $m_a$ and $u$ solutions of the same magnitude as 
those obtained from $u_o$. 
In particular, $u_+$ or $u_-$ may sometimes 
lack a valid solution when $m_b \approx m_c$,   
and instead require that $m_a$ become complex. 

Both $u_+$ and $u_-$ do, however, provide some additional
classes of mass ratio solutions that $u_o$ does not allow.
(See Appendix B.) 
In all but one of these additional classes   
{\it very} fine tuning of $m_a$ and $u$ is required
to produce a specific three quark mass ratio.
The non--fine tuning exception is a $u_+$ class of   
{\it extremely}--weak coupling solutions 
with $m_a\gg \Lambda$ and $u\gg \Lambda^2$.  

To conclude this section, we comment that the corresponding
Hanany and Oz family of hyperelliptic curves for $N_c=2$, $N_f=3$ is, 
\beqn
y^2 = \left(
      x^2 - u + \Lambda (\frac{m_a + m_b + m_c}{8} + \frac{x}{4}) \right)^2 
     - \Lambda^3 \prod^c_{i= a} \left(x + m_i\right)\, .
\label{hza2} 
\eeqn
Hanany and Oz have also given the curves for $N_c=3$, $N_f=3$,
\beqn
y^2 = 
\left(x^3 - u_2 x - {u_3 \over 3} + {\Lambda^3 \over 4}\right)^2 
     - \Lambda^3 \prod^c_{i= a} \left(x + m_i\right)\, .
\label{hza3}
\eeqn
(See Eqs.\ (5.5) and (4.13) of \cite{ho}, respectively.)

\section{Discussion}

We have studied bare quark mass ratios at the singular points on the
complex $u$--plane of Seiberg--Witten $N=2$ supersymmetric $N_f=2$ 
and $N_f=3$ $SU(N_{c}=2)$ theory.
We have shown that large bare mass hierarchies at the singular points 
can occur for both the two quark and three quark models.
For $N_f=2$ we found that demanding large bare mass ratios
at singularities placed the singularities in the strong coupling region.
In contrast, for $N_f=3$ we determined the respective singularities 
could be located in either the strong or weak coupling regions.

We would emphasize that in general the bare masses are not the physical
masses; only in the weak coupling limit do the bare masses become physical masses.
Nonetheless, large bare quark mass hierarchies 
at the singularities of the $N=2$ $SU(N_{c}=2)$ parameter space 
for two or three quark flavors may suggest a possible explanation for the 
phenomenologically known three generation mass hierarchy. 
Such an explanation would need not  
depend on non--renormalizable terms in the superpotential.
This explanation would require an extrapolation from the $N=2$, $N_{c}=2$ theory
discussed herein to the $N=1$, $N_{c}=3$ case. This suggests that 
the $N=2$, $N_{c}=3$ case should be investigated as a next step. This is, however, 
beyond the scope of this letter and so we leave this for future research.      

It is interesting to note that the Seiberg--Witten equation for the family of
curves can be obtained from M--theory as shown by Witten \cite{ew}.  
Additional relevant information is available in 
Ennes, et al.\ \cite{enn} and the references cited therein.
Witten studied the $N=2$ supersymmetric gauge theories in four dimensions by
formulating them as the quantum field theories derived from a configuration of
various D--branes.  
He considered, for example, $N_c = N_f = 3$ ($c$ is color, $f$
is flavor) quantum field theory of two parallel five--branes connected by 3
four--branes, with 3 six--branes between them in Type IIA superstring theory on
$R^{10}$, and reinterpreted this configuration in M--theory.  World volumes of
five--branes, four--branes, and six--branes are parametrized by the coordinates
$x^0x^1x^2x^3$ and $x^4x^5$, $x^0x^1x^2x^3$ and $x^6$, $x^0 x^1x^2x^3$ and
$x^7x^8x^9$, respectively.

In M--theory, the above brane configuration can be reinterpreted as a
configuration of a single five--brane with world volume $R^4 \times \sum$ where
$\sum$ is the Seiberg--Witten curve.  It yields the structure of the Coulomb
branch of the $N=2$ theory.  The curve $\sum$ is given by an algebraic equation
in $(x, y)$ space where $x = x^4 + i x^5$ and 
$y = \exp [ - ( x^6 + i x^{10})/R]$. In terms of $\tilde y = y + B/2$, one 
obtains
\beqn
\tilde y^2 = ( B(x)^2 /4) - \Lambda^3 \prod^c_{i =a} (x + m_i),
\label{ewa}
\eeqn
where $B(x) = e (x^3 + u_2 x + u_3)$.
This is the hyperelliptic curve for $N_c= N_f=3$ obtained by
Hanany and Oz in (\ref{hza3}).
 
\section{Acknowledgments}

This work is supported in part by DOE Grant DE--FG--0395ER40917 (GC).
\newpage
\appendix

\section{$\mbf N_c=2$, $\mbf N_f=3$ Discriminant Boundary Constraint As 
Function of Bare Masses, $\mbf m_a$, $\mbf m_b$, $\mbf m_c$ 
and Complex Parameter $\mbf u$} 

\beqn
\Delta_{BNDY} &=& \phantom{+}
t^8 [ (m_a^4 + m_b^4 + m_c^4) - 2 (m_a^2 m_b^2 + m_a^2 m_c^2 + 2 m_b^2 m_c^2 ) + 3 m_a^{4/3} m_b^{4/3} m_c^{4/3} ]\nolabel\\ 
&&+t^7 [  -4 (m_a^3 m_b m_c +m_a m_b^3 m_c +m_a m_b m_c^3) + 12 m_a^{5/3} m_b^{5/3} m_c^{5/3} ]\nolabel\\  
&&+t^6 [   4 (m_a^6 + m_b^6 + m_c^6) - 6 (m_a^4 m_b^2 + m_a^2 m_b^4 + m_a^4 m_c^2 + m_b^4 m_c^2 + m_a^2 m_c^4 + m_b^2 m_c^4 )\nolabel\\ 
&&\phantom{+t^6}
       +24 m_a^2 m_b^2 m_c^2  - 4 (m_a^4  + m_b^4  + m_c^4 ) u +16 (m_a^2 m_b^2 + m_a^2 m_c^2  +  m_b^2 m_c^2 ) u \nolabel\\
&&\phantom{+t^6}
       -36 m_a^{4/3} m_b^{4/3} m_c^{4/3} u - 2 ( m_a^2 + m_b^2 +  m_c^2 ) u^2 +  6 m_a^{2/3} m_b^{2/3} m_c^{2/3} u^2 ]\nolabel\\
&&+t^5 [-24 ( m_a^5 m_b m_c   +m_a m_b^5 m_c   +m_a m_b m_c^5 ) -12 ( m_a^3 m_b^3 m_c + m_a^3 m_b m_c^3 +m_a m_b^3 m_c^3)\nolabel\\ 
&&\phantom{+t^5}
     +108 m_a^{7/3} m_b^{7/3} m_c^{7/3} + 52 (m_a^3 m_b m_c +m_a m_b^3 m_c +m_a m_b m_c^3 ) u \nolabel\\
&&\phantom{+t^5}
       - 156 m_a^{5/3} m_b^{5/3} m_c^{5/3} u ]
\nolabel\\ 
&&+t^4 [-27 (m_a^4 m_b^4 + m_a^4 m_c^4 + m_b^4 m_c^4) +99 m_a^{8/3} m_b^{8/3} m_c^{8/3}\nolabel\\ 
&&\phantom{+t^4}
     - 6 (m_a^4 m_b^2 m_c^2 + m_a^2 m_b^4 m_c^2 + m_a^2 m_b^2 m_c^4)
\nolabel\\
&&\phantom{+t^4}
    +36 (m_a^4 m_b^2 + m_a^2 m_b^4 + m_a^4 m_c^2 + m_a^2 m_c^4 
    + m_b^4 m_c^2 + m_b^2 m_c^4) u 
\nolabel\\
&&\phantom{+t^4}
    -216  m_a^2 m_b^2 m_c^2 u  - 8 (m_a^4 + m_b^4 + m_c^4) u^2 
    - 46 (m_a^2 m_b^2 + m_a^2 m_c^2  +  m_b^2 m_c^2) u^2 
\nolabel\\ 
&&\phantom{+t^4}
+162 m_a^{4/3} m_b^{4/3} m_c^{4/3} u^2 + 8 (m_a^2 + m_b^2 + m_c^2)u^3
    - 24 m_a^{2/3} m_b^{2/3} m_c^{2/3} u^3 ]\nolabel\\
&&+t^3 [  36 (m_a^3 m_b^3 m_c + m_a^3 m_b m_c^3 +m_a m_b^3 m_c^3 )u  -108  m_a^{7/3} m_b^{7/3} m_c^{7/3} u
\nolabel\\ 
&&\phantom{+t^3}
   - 40 (m_a^3 m_b m_c +m_a m_b^3 m_c +m_a m_b m_c^3) u^2+120 m_a^{5/3} m_b^{5/3} m_c^{5/3} u^2 ]
\nolabel\\ 
&&+t^2 [ -4 (m_a^2 m_b^2 + m_a^2 m_c^2 + m_b^2 m_c^2 ) u^3 +12  m_a^{4/3} m_b^{4/3} m_c^{4/3} u^3
\nolabel\\ 
&&\phantom{+t^2}
   + 4 (m_a^2 + m_b^2 + m_c^2)u^4 - 12 m_a^{2/3} m_b^{2/3} m_c^{2/3} u^4 ]
\label{bdabct}
\eeqn
\hfill\vfill\eject

\section{Bare Mass Ratios and Related $\mbf u_o$ Singularity Locations}

\def\hbi{\hbox to .3truecm{\hfill}}
\def\la{\Lambda}
\def\tp#1{\times 10^{#1}}

\begin{center}
\begin{tabular}{|l||rr|}
\hline 
\hline
Mass Ratio       &  $u_o$ &       \\
$\frac{m_a}{m_a}:\frac{m_b}{m_a}:\frac{m_c}{m_a}$    
                 &  $m_a$ $(\la)$ & $u$  $(\la^2)$\\ 
\hline
\hline
$1:.007:.00002$  &  1600  &  74  \\
$1:.008:.00003$  &  1400  &  73  \\
$1:.5:.5$        &   .32  & .064 \\
$1:.1:.1$        &  2.7   & .42  \\
$1:.01:.01$      &   58   & 7.6  \\
$1:.001:.001$    &  1300  & 160  \\
\hline
\hline
\end{tabular}
\end{center}
Table I. Bare Mass Quark Ratios and the Related $u_o$ Singularity. 
Listed here are the $m_a$ and $u$ solutions of 
the $u_o$ singularity for a few quark bare mass ratios.
In these examples 
$0.1\, \Lambda   \lsim m_a \lsim 2000 \Lambda$ and
$0.1\, \Lambda^2 \lsim u   \lsim 200  \Lambda^2$.
For a given mass ratio, $u_{-}$ and $u_{+}$ generally 
offer $m_a$ and $u$ solutions similar in magnitude to those of
$u_o$. (For the $u_{-}$ solutions $m_a$ and $u$ are of opposite sign.)
However, corresponding 
$u_{-}$ or $u_{+}$ solutions sometimes fail to exist when $m_b \approx m_c$.

For generic mass ratios 
(excepting those where $m_b/m_a\sim m_c/m_a \sim {\cal{O}}(1)$), 
the $u_{-}$ and $u_{+}$ singularities offer
three additional classes of strong coupling solutions 
that $u_o$ does not provide. 
These solutions involve fine tuning of $m_a$ and $u$ though.
The $u_+$ singularity yields solutions where (i) 
${\cal{O}}(m_a)\sim \frac{1+ \beta}{32}\Lambda$, 
with $|\beta|\lt .01$ and 
$\frac{1}{256} \Lambda^2 \le u< \frac{1}{160}\Lambda^2$, and
(ii) ${\cal{O}}(m_a)\lsim 10^{-21}\Lambda$ 
and $u= \frac{1+\eps}{256} \Lambda^2$, with $|\epsilon|< 10^{-8}$.  
The $u_-$ singularity solutions produce
the same mass ratios and  
${\cal{O}}(m_a)\lsim 10^{-21}\Lambda$ mass as in (ii) above,  
but require ${\cal{O}}(u)\lsim 10^{-19} \Lambda^2$ instead.

The $u_+$ singularity also offers {\it extremely}--weak coupling solutions 
where ${\cal{O}}(m_a)\sim 10^{8\,\, {\rm to}\,\, 12} \Lambda$ and   
${\cal{O}}(u)\sim 10^{9\,\, {\rm to}\,\, 14} \Lambda^2$. 
For example, the $1:.007:.00003$ ratio occurs at
$m_a\app 6.2\times 10^{11}\Lambda$ and 
$u\app 1.8\times 10^{14}\Lambda^2$.

\def\ATMP#1#2#3{{\it Adv.\ Theor.\ Math.\ Phys.}\/ {\bf#1} (#2) #3}
\def\AP#1#2#3{{\it Ann.\ Phys.}\/ {\bf#1} (#2) #3}
\def\NPB#1#2#3{{\it Nucl.\ Phys.}\/ {\bf B#1} (#2) #3}
\def\NPBPS#1#2#3{{\it Nucl.\ Phys.}\/ {{\bf B} (Proc. Suppl.) {\bf #1}} (#2) 
 #3}
\def\PLB#1#2#3{{\it Phys.\ Lett.}\/ {\bf B#1} (#2) #3}
\def\PRD#1#2#3{{\it Phys.\ Rev.}\/ {\bf D#1} (#2) #3}
\def\PRL#1#2#3{{\it Phys.\ Rev.\ Lett.}\/ {\bf #1} (#2) #3}
\def\PRT#1#2#3{{\it Phys.\ Rep.}\/ {\bf#1} (#2) #3}
\def\PTP#1#2#3{{\it Prog.\ Theo.\ Phys.}\/ {\bf#1} (#2) #3}
\def\MODA#1#2#3{{\it Mod.\ Phys.\ Lett.}\/ {\bf A#1} (#2) #3}
\def\IJMP#1#2#3{{\it Int.\ J.\ Mod.\ Phys.}\/ {\bf A#1} (#2) #3}
\def\nuvc#1#2#3{{\it Nuovo Cimento}\/ {\bf #1A} (#2) #3}
\def\RPP#1#2#3{{\it Rept.\ Prog.\ Phys.}\/ {\bf #1} (#2) #3}

               
\vfill\eject

\bigskip
\medskip

\def\bibiteml#1#2{ }
\bibliographystyle{unsrt}

\hfill\vfill\eject
\end{document}